# Dissociative electron attachment to gold(i) based compounds: 4,5-dichloro – 1,3-diethyl – imidazolylidene trifluoromethyl gold(i)

[1]Maria Pintea, [1]Nigel Mason, [1]Anna Peiró – Franch, [1]Ewan Clark, [1]Kushal Samantha, [2]Cristiano Glessi, [2]Inga Lena Schmidtke, [3]Thomas Luxford

[1]School of Physical Sciences, University of Kent, CT2 7NZ, Canterbury, UK
[2]Department of Chemistry, University of Oslo, Oslo, Norway
[3]Department of Chemistry, J. Heyrovský Institute of Physical Chemistry of the Czech Academy of Sciences, Czech Republic

**Abstract**

With the use of proton-NMR and powder XRD (XRPD) studies, the suitability of specific Au FEBID precursors has been investigated to low electron energy, structure, excited states and resonances, structural crystal modifications, flexibility and vaporization level. Uniquely designed precursor to meet the needs of focused electron beam induced deposition (FEBID) at the nanostructure level, the 4,5-dichloro – 1,3-diethyl – imidazolylidene trifluoromethyl gold(i) is a compound that proves its capability to create high purity structures, and its growing importance between other $AuIm_x$ and $AuCl_nB$ (where x, n are the number of radicals, B = CH, $CH_3$ or Br) compounds in the radiation cancer therapy increases the efforts to design more suitable bonds in processes of SEM deposition and in gas-phase studies. The investigation done to its powder shape using the XRPD XPERT[3] Panalytical diffractometer based on $CoK_α$ lines shows changes to its structure with temperature, level of vacuum and light; the sensitivity of this compound making it highly interesting to the radiation research in particular. Used for FEBID, through its smaller number of C, H and O atoms has lower levels of C contamination in the structures and on the surface, but it replaces these bonds with C – Cl and C – N bonds that have a lower bond breaking energy, but still needing an extra purification step in the deposition process, whether is $H_2O$, $O_2$ or H jets.

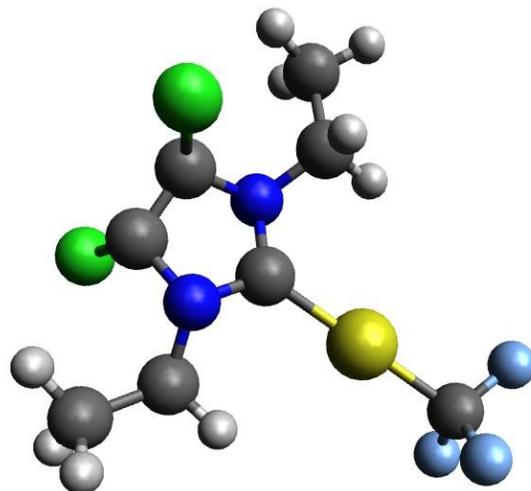

Fig 1. 4,5-dichloro – 1,3-diethyl – imidazolylidene trifluoromethyl gold(i) (Au yellow, N blue, H white, C grey, F light blue, Cl green)

**Introduction**

The increased importance in the pharmaceutical drug development and cancer studies, the gold(i) compounds are developed from synthesis step to quantum simulations and molecular dynamics analysis, as in our case, or for catalysed reactions and reductive elimination/migratory insertion reactions [1]. The most worth mentioning application of the gold(i) compounds is the inhibition of bacteria such as Escherichia Coli [6], through bonding of the DHFR to the C, F or P of the gold compound, where a high reduction in the level of DHFR in solution with gold(i) compared to non-gold(i) is observed. A level of reduction of 0.1 from 2.2 to 2.1 of denaturated DHFR and 1.2 to 1 of the native DHFR is observed with inhibitory constants of 2.25μM, 1.1μM and 8.63μM for 4-benzoic acid-diphenyl-phosphene gold(i) chloride, 2-benzoic acid-diphenyl-phosphane gold(i)-chloride and 4,5-dichloroimidazolato-N-triphenylphosphine–gold(i) extending the use of the gold(i) compounds to the treatment of inflammatory infections, pneumonia, E. Coli and cancer.

Through the use of velocity map imaging technique and DEA mass spectroscopy studies, employed for multiple analysis involving Au compounds or compounds on gold substrates, we determine the fragmentation pathways with implications to focused electron beam deposition. At 157nm, the velocity map imaging study of diatomic gold in combination with DFT and ab initio calculations brings insight to the dynamics of the Au – Au vibrational and excitations modes, bonding between species with d-electrons valence, as well as the branching ratios for Au $5d^9 6s^2 (^2D_{3/2})$ and Au $5d^9 6s^2 (^2D_{5/2})$ *[1]*. The optical absorption spectrums of Au in vapour form shows the allowed transition states between 211 - 229nm from $^1\Pi_u(II)$ to $X^1\Sigma_g^+$, and isolates two dissociation processes, first one at a photon energy of 2.301 – 2.311eV for Au $5d^{10} 6s^2 (^2D_{5/2})$ and the second one at 3.437 – 3.447eV for Au $5d^{10} 6s^2 (^2S_{1/2})$ + Au $5d^9 6s^2 (^2D_{5/2})$, showing particularity for gold cluster processes and the presence of the 6s orbitals combined with the relativistic effects of the s electrons. For nanotechnology applications, the assisted deposition of Au compounds has been done successfully by *Schawrav et al [28]* with $H_2O$ as oxidative enhancer resulting pure Au nanostructures with a resistivity of 8.8μΩcm and 91 at.% purity of the structure. The Au content of the nanostructures resulting from the focused electron beam induced deposition of $Me_2Au(tfac)$ was improved to reach values of 72 at.% through the refining of the electron beam parameters, and further to hit high purity levels of ~90 at.% through the plasma assisted structure post-processing in *Ref. [29]*. *Chien et al [30]* reports carbon content up to 60% in their Au deposited nanostructures through their newly developed localized surface plasmon resonance measurement (built to enhance structure content reading) and a reduction down to 20% of the carbon content through the $H_2O$ treatment of the nanostructures.

In the normal non-assisted deposition of $CF_3$ - Au containing compounds *[31], [32]* values of Au content in the deposits of 22at% in the case of $CF_3AuCNMe$ and 14at% for $CF_3AuCNBu$ *[32]* were obtained with values of decomposition and sublimation temperatures of the two compounds evaluated of 51°C and 80°C ($CF_3AuCNMe$) and, 39°C and 126°C ($CF_3AuCNBu$) respectively. $CF_3$ – Au containing precursors are known to have very good sublimation and decomposition temperatures becoming highly sought precursors for FEBID deposition, though the lower levels of Au content and high C contamination (>60at%) are indications of the need of a post-processing treatment or assisted deposition. The $Me_2Au(Acac)$ presents comparable results when deposited and annealed at 100 - 300°C forming structures close to 14nm *[33]*, but at the same time reducing the carbon content at 300°C under $H_2$ jet to almost 0 at% and removing it out of the lattice through heating. *Gruber et al [34]* report the growth of $AuC_x$ nanopillars results of the FEBID of $Me_2Au(acac)$ with a height of 2 μm for the development of the 3D plasmonic gold nanoantennae, as one of the many applications of the induced chemistry at the nanoscale. The focus is indeed on the composition of the nanopillars that are further annealed (to 300°C) and purified using $H_2O$ jets at room temperature. The growth of nanoantennae and nanopillars have put the basis of a new lithographic method based on FEBIP using cooling to lower than 0°C of the substrate and thin films and further irradiated using e-beams to form structures [35], or, more sophisticated methods as GIS (gas injection systems) and computer assisted deposition for the creation of highly complex and accurate 3D nanostructures [36]. The same methods have been applied to growing carbon nanotubes [37], carbene nanostructures [38] and cold ice organic nanostructures [35]. Nanostructures have been printed in [39] using Au based compounds in reactive atmospheres [40] with very successful outcomes.

**Experimental Section**

**VsMI/Mass Spectroscopy.** The experimental equipment consists of a high-vacuum chamber with pressures in the range of $10^{-9}$mbar helped by a SCROLLVAC SC5D scroll pump and a gas-line oil pump Edwards RV3 with pressures in the range of $10^{-3}$mbar. An electron gun is mounted on the top flange of the chamber intersecting at 90° the molecular beam and in-line to the electron gun, a three-plate Chevron pattern MCP detector. A puller, pusher and flight tube assembly are connected to the detector for guiding the negatively charged ions to the phosphor screen. A CCD industrial camera is used for capturing the ions accelerated to different velocities. A pair of Helmholtz coils is placed on the top and bottom of the chamber with the purpose of creating a magnetic field with values up to 50Gauss that controls the guide path of the particles (ions/molecular fragments, electrons). The simple assembly: electron gun, the detector assembly, the flight tube guiding the ions to the phosphor screen and a CCD camera for capturing the negative fragments, is helped by a Behlke 100ns to 25ns slicer, that would physically select a certain time length slice of the Newton sphere of ions that would further be imaged by the camera and detected by the MCP data acquisition modules for mass discrimination. Similar set-ups [8], [9], [10] have been used at Tata Institute, India for imaging of negative ions.

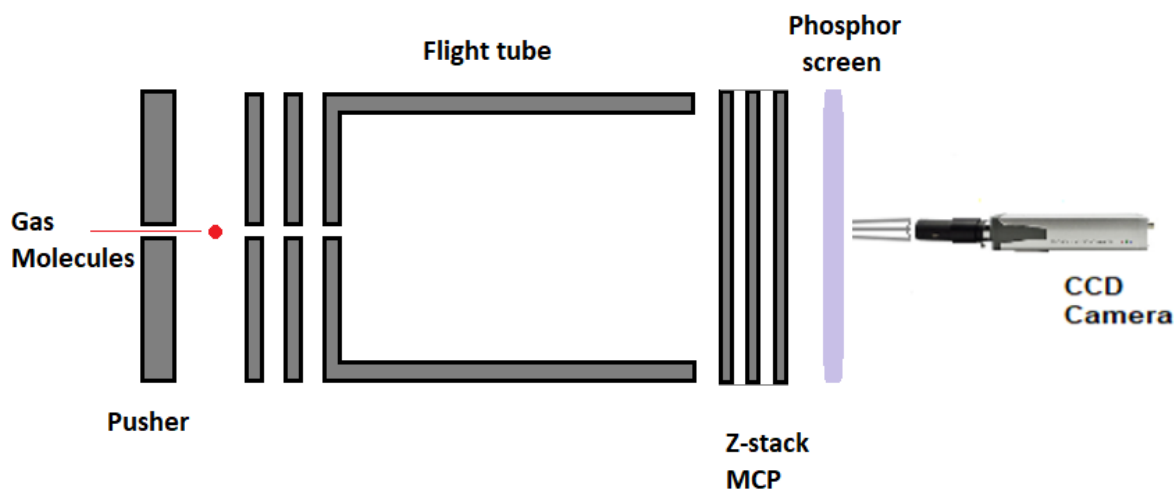

Fig 2. MCP VsMI detector assembly

The increase in the number of detector's plates, reduces the aberration of the equipment, and improves the energy range. The VsMI in the **Fig 2** assembly uses an energy range of 0 - 50eV specifically for low electron energy applications. The phosphor screen is employing a thin tungsten with >98% transmission rate foil mounted on a brass ring. The detection of the negative ions is calibrated against a system of two molecules, $O_2$ at 6.5eV and $SO_2$. The kinetic energies spectrum and angular distribution follow the same rules and should be less than 0.1% to the $O_2$ and $SO_2$ spectrums.

**XRD (X-ray Powder Diffraction).** To determine the structural characteristics of the crystalline sample XRD measurements were employed using a XPERT$^3$ Panalytical diffractometer based on CoK$_\alpha$ with a time step of 150s/step and a step size of 0.0167. The angle of diffraction is 5 – 80deg at a rate of 40kV and 40mA. The measurements were acquired over the duration of $1^{1/2}$ hours. The diffraction specific wavelength is set to a value of 1.5406Å.

**NMR (Nuclear Magnetic Resonance).** The proton NMR data acquisition was done using a JEOL ECS 400MHz NMR spectrometer at 25°C with a sensitivity of 280 (0.1% ethyl benzene) for $^1$H and $^{19}$F, with an automatic Bruker SampleXpress sample charger run by a 500MHz electric DC motor having a 60 sample carousel controlled by ICON-NMR software and equipped with barcode reader registration, with the samples kept at a temperature between 5 - 30°C and a separate cryo-fit mounting kit for sample cooling. The sample charger and sample unit were both controlled by the Bruker Avance III 400MHz controller unit.

**Results and Discussion**

**Structure characterization.** The 4,5-dichloro – 1,3-diethyl – imidazolylidene trifluoromethyl gold(i) is a gold compound synthesized by the Chemistry Department of the University of Oslo. The linear formula of the compound is $C_8H_{10}Cl_2N_2AuF_3$, and it has a mass of 459.05amu. The schematic of the compound is presented in **Fig 3**.

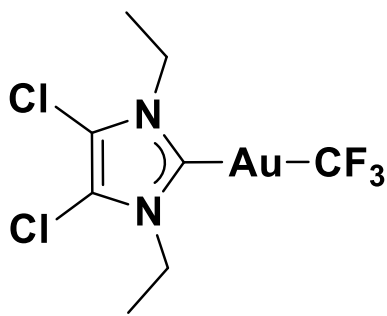

Fig 3. Schematics of 4,5-dichloro – 1,3-diethyl – imidazolylidene trifluoromethyl gold(i)

A very important characteristic to our molecular dynamic simulations and cross-checking of our experimental results is the bond distance to C, Cl, N and H and bond angles to C, Cl, N and H. Multiple sources on trifluoromethyl gold(i) *[2], [3], [4]*, present the bond distances for the gold(i) compounds as 0.05Å higher than

gold(ii) compounds. In the study of *Gil-Rubio and Vicente [1]*, the bond distances for the most common gold(i) compounds are experimentally determined with values in the range of ~2.04Å presented in **Table 1**.

| Compound | d(Au – C) /Å, X = F |
|---|---|
| Ph$_3$P – Au - CX$_3$ | 2.045 |
| IPr – Au - CX$_3$ | 2.042; 2.030 |

Table 1. L – Au – CF$_3$ bond distances *[1]*

The characteristics of the trifluoromethyl complexes come into a higher bond distance MC – F than C – F, as well as a decrease of F – C – F bond angle and increase in the M – C – F bond angle to the tetrahedral symmetry point group. The bond distance Au – CF$_3$ is shorter than the Au – CH$_3$, and the Au – C / Au – Cl, where Au – Cl bond distance is in the range of ~ 2.27Å *[3]* for [AuCl]$^-$, similar in scattering processes to Au(iii) anion [AuCl$_4$]$^-$, vibrational bands are weak Au 2p$_{3/2}$ to 5d transitions, the so called white line, for [AuCl$_2$]$^-$ ions *[3]* and this weak transitions are the result of a transition from Au 6s/5d hybrid partially occupied to the highest energy level occupied HOMO orbital. For our standard compound we obtain the HOMO and LUMO orbital as orbitals 76 and 77, while the total SCF density would contain a number of 683 occupied and unoccupied orbitals. Both the HOMO and the LUMO of 4,5-dichloro – 1,3-diethyl – imidazolylidene trifluoromethyl gold(i) are presented in **Fig 4**.

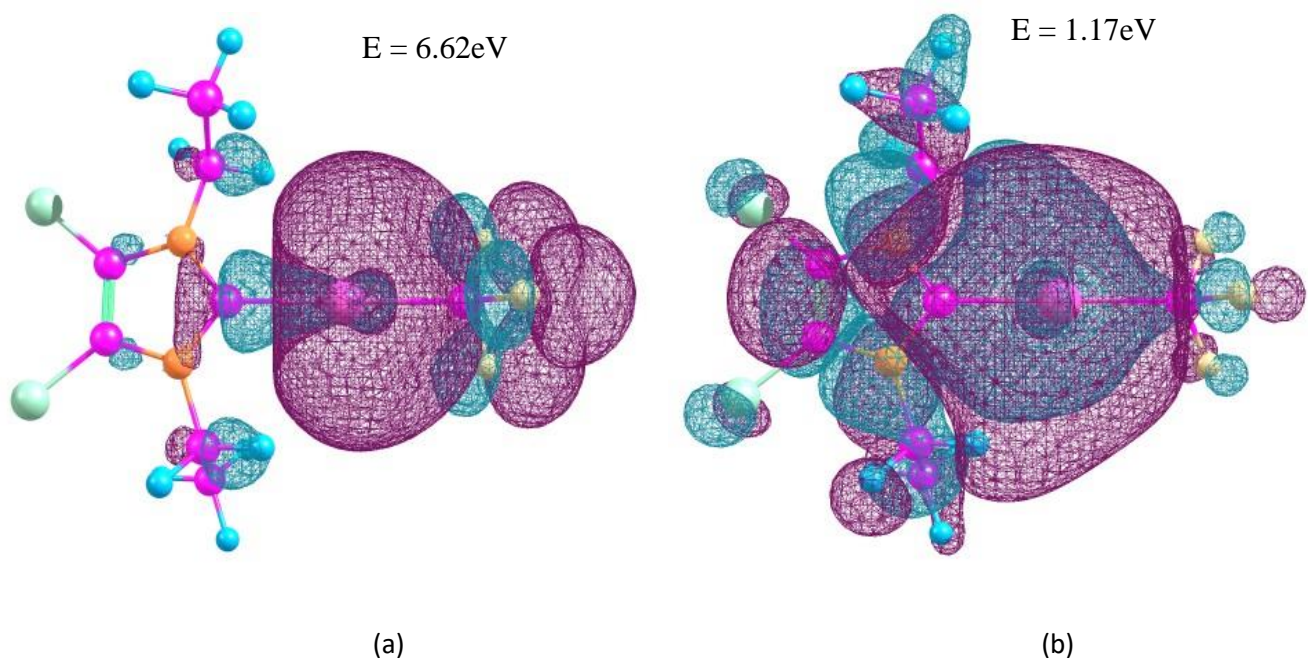

(a)          (b)

Fig 4. HOMO/LUMO orbitals of 4,5-dichloro – 1,3-diethyl – imidazolylidene trifluoromethyl gold(i) (F yellow, C pink, H blue, N orange, Cl green, Au magenta) : (a) HOMO (b) LUMO

The bond distances calculated using B3LYP/Def2TZVPP are longer than the free methyl radical bond distances, and more imbalanced ranging from 1.087Å to 2.076Å. The three C atoms of the methyl radical have the bond lengths of 1.089/1.090Å, equally spaced in all directions with an angle <HCH of 108.1deg. The angles characteristic to the methyl radicals in our compound are 106.02deg (<HCH) to 112.71deg (<CCH). The bond lengths and angles of 4,5-dichloro – 1,3-diethyl – imidazolylidene trifluoromethyl gold(i) are presented in **Table 2**. We report a bond distance of C – Au of 2.065Å from C7 - Au8 and a second bond distance to $CF_3$, Au8 – C9 of 2.076Å, the affinity to $CF_3$ being higher than to $CN_2$ in our case. Lower bond distances of Au – C ligands have been reported in *[5]* with values declared for the Au – $CF_3$ bond of all gold(i) trifluoromethyl complexes to be between 2.031Å and 2.046Å. Both the bond distances to $CN_2$, C3 – N5 and C4 – N6 have values of 1.384Å and angles to axial plane of 125.3deg, while the C3 – Cl1 and C4 – Cl2 bond distances are 1.702Å set at equally spaced angles of 129deg. All C – F bond lengths (**Table 2**) from $CF_3$ have values of 1.372Å balanced with an angle of 104.6deg, longer than in the free trifluoromethyl radical with distances of the C – F bond of 1.318Å and an angle of 110.76deg, 6.16deg lower than in our calculations. The simple ethyl radicals C13 – H25 and C13 – H19 have bond lengths of 1.087Å, and 1.089Å, respectively.

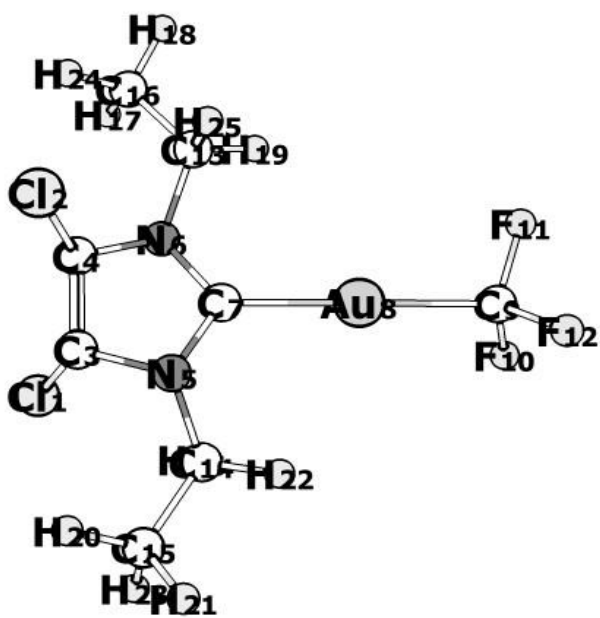

| Type of bond | Bond lengths (Å) |
|---|---|
| Au8 – C7 | 2.065 |
| Au8 – C9 | 2.076 |
| C7 – N5 | 1.356 |
| N5 – C3 | 1.384 |
| C3 – Cl1 | 1.702 |
| C3 – C4 | 1.357 |
| N5 – C14 | 1.469 |
| C13 – C16 | 1.523 |
| C13 – H19 | 1.087 |
| C13 – H25, C16 – H18 | 1.089 |
| C16 – H17/24 | 1.090 |
| C9 – F10 | 1.372 |

Table 2. Bond lengths and angles from B3LYP/Def2TZVPP calculations

The highest values of the Au8 – C7 bonds are for the B3LYP/CEP-121 basis set of 2.066Å, 0.001Å higher than the calculations at B3LYP/Def2TZVPP level of theory, 0.002Å higher than the calculations at B3LYP/QZVP and 0.014Å; the highest discrepancy is obtained using SDD basis set with the lowest bond length value of 2.052Å. The 4,5-dichloro – 1,3-diethyl – imidazolylidene trifluoromethyl gold(i) compound has the Au – C and C – Au bond lengths 0.4Å higher than the declared value for the Au – C bond in $AuIm_2$ in *[65], [66]* of 1.7Å - 2.06Å by *Liu, Xiong, Diem Dau et al (2013)* and *Benitez et al (2009)*.

The accuracy of MP2 methods compared to the one of B3LYP and HF is very low for very complex molecules containing a high number of atoms or organic parts (peptides, alanine) *[64]*. *Kaminsky et al (2008) [64]* calculates the error of the MP2 methods with the basis set as being 20 to 30 kJ/mol in the electronic energy calculations; the values we report for the MP2 with QZVP basis set are the shortest distances Au8 – C7 to the $Cl_2$-phenyl ring; the calculations, results of Def2TZVPP have values higher with 0.007Å of the Au8 – C7 distance. All MP2 level calculations for all basis-sets have shorter bond length values, the MP2/SDD Au8 – C7 has a value of 2.038Å, while the MP2/CEP-121G has a value slightly higher of 2.045Å, but still lower than B3LYP calculations with the same basis set with values of 2.052Å and 2.066Å, respectively.

| | | | | | | | | | | | | |
|---|---|---|---|---|---|---|---|---|---|---|---|---|
| | | | | | | | | | | | Distances in Å | |
| Basis Set/ Method Bond * | Def2TZVPP | | | SDD | | | CEP-121G | | | QZVP | | |
| | B3LYP | MP2 | HF | B3LYP | MP2 | HF | B3LYP | MP2 | HP | B3LYP | MP2 | HF |
| Au8 – C7 | 2.065 | 1.990 | 2.108 | 2.052 | 2.038 | 2.097 | 2.066 | 2.045 | 2.066 | 2.064 | 1.983 | 2.106 |
| Au8 – C9 | 2.076 | 2.025 | 2.098 | 2.071 | 2.071 | 2.105 | 2.079 | 2.073 | 2.079 | 2.076 | 2.019 | 2.099 |
| C7 – N5 | 1.356 | 1.359 | 1.335 | 1.380 | 1.396 | 1.356 | 1.388 | 1.402 | 1.388 | 1.355 | 1.357 | 1.335 |
| N5 – C3 | 1.384 | 1.371 | 1.378 | 1.400 | 1.407 | 1.393 | 1.410 | 1.411 | 1.410 | 1.383 | 1.368 | 1.378 |
| C3 – Cl1 | 1.702 | 1.688 | 1.696 | 1.771 | 1.791 | 1.756 | 1.781 | 1.793 | 1.781 | 1.701 | 1.685 | 1.694 |
| C3 – C4 | 1.357 | 1.375 | 1.330 | 1.370 | 1.396 | 1.338 | 1.376 | 1.401 | 1.376 | 1.357 | 1.373 | 1.329 |
| N5 – C14 | 1.469 | 1.462 | 1.464 | 1.485 | 1.501 | 1.481 | 1.496 | 1.505 | 1.496 | 1.469 | 1.460 | 1.464 |
| C13 – C16 | 1.523 | 1.517 | 1.520 | 1.538 | 1.554 | 1.531 | 1.545 | 1.555 | 1.545 | 1.523 | 1.515 | 1.520 |
| C13 – H19 | 1.087 | 1.087 | 1.077 | 1.093 | 1.100 | 1.078 | 1.093 | 1.098 | 1.093 | 1.087 | 1.087 | 1.077 |
| C13 – H25, C16 – H18 | 1.089 | 1.088 | 1.079 | 1.095 | 1.102 | 1.079 | 1.095 | 1.100 | 1.095 | 1.088 | 1.087 | 1.079 |
| C16 – H17/24 | 1.090 | 1.088 | 1.083 | 1.096 | 1.104 | 1.083 | 1.096 | 1.102 | 1.096 | 1.089 | 1.087 | 1.082 |
| C9 – F10 | 1.372 | 1.366 | 1.342 | 1.430 | 1.454 | 1.401 | 1.438 | 1.458 | 1.438 | 1.371 | 1.364 | 1.342 |

Table 3. Bond lengths of $ImEtAuCl_2CF_3$ optimized at multiple levels of theory

The chlorine atoms to the phenyl ring bond lengths range from 1.685Å calculated at MP2/QZVP to 1.793Å at MP2/CEP-121G, both Cl atoms being placed symmetrically at a 129.22deg angle to the C3 and C4 atoms of the phenyl ring. Overall the structure is balanced at the central symmetry plane formed by the two C atoms and the Au, C7 – Au8 – C9, while a 110.83deg to C (in $CH_2$) and 90deg to the symmetry plane is obtained for the two ethyl – methyl groups.

## XRD measurements of 4,5-dichloro – 1,3-diethyl – imidazolylidene trifluoromethyl gold(i) for nanomaterials characterization

**Crystal structure analysis.** The $ImEtAuCl_2CF_3$ crystal simulation was done in Vesta software on a preoptimized molecular structure in Gaussian 16 at B3LYP/Def2TZVPP level of theory using density functional theory calculations with full orbital populations and triple-ζ frequency calculations. A number of bond lengths and the unit cell have been optimized by trial and error to obtain the slab of the crystal with different γ-orientations of the crystal planes.

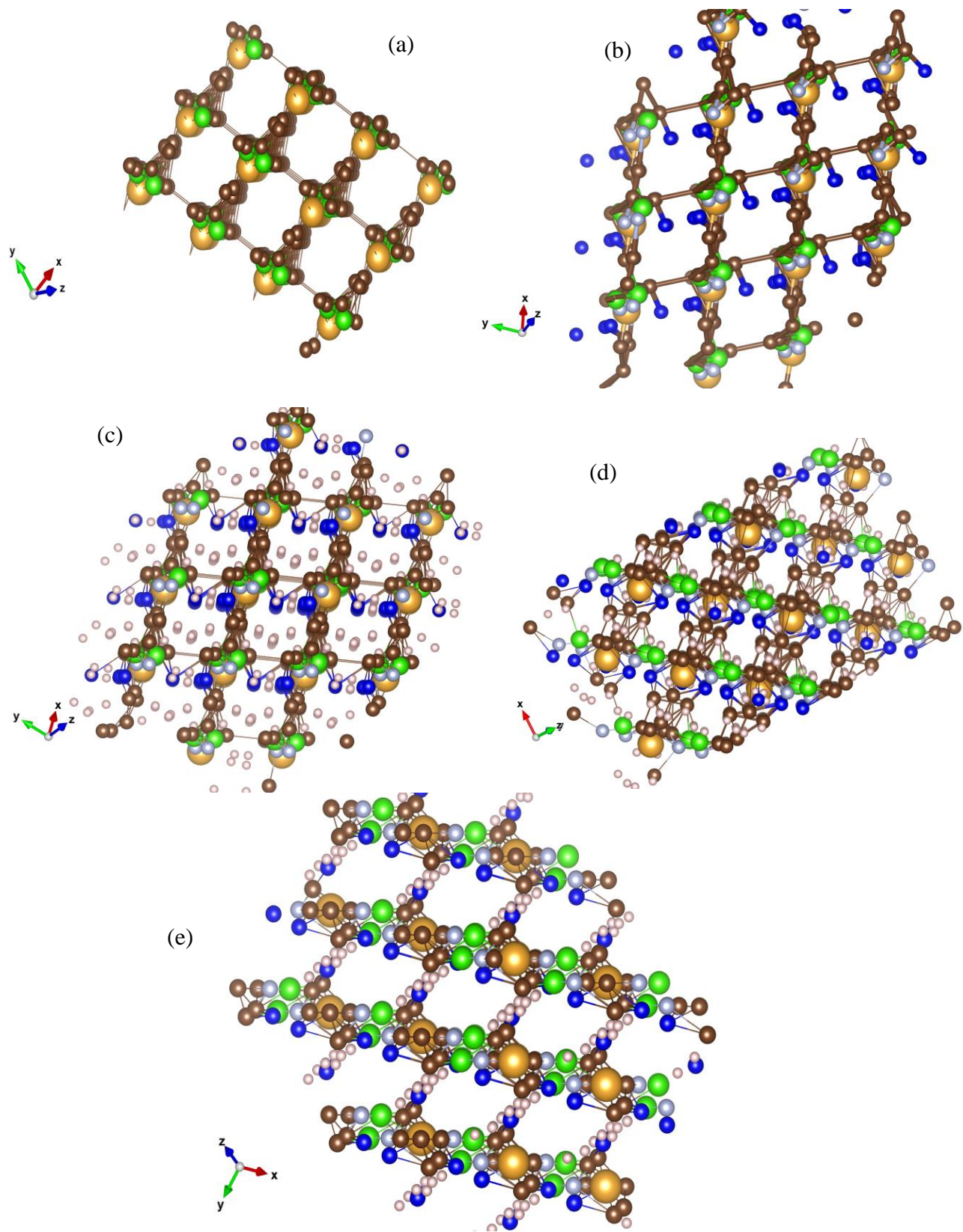

Fig 5. Crystal structure simulations of 4,5-dichloro – 1,3-diethyl – imidazolylidene trifluoromethyl gold(i) for nanomaterials characterization. Characteristic to the ImEtAuCl$_2$CF$_3$ crystal matrix is the bonding at 60deg of the carbon basket surrounding the Au atom to the next C-ring structure through a fluorine - hydrogenated C bond, while the nitrogen would have a higher noble metal affinity bonding to the Au – C structure forming AuNCl$^-$ sites with high electron affinity compared to the C lattice surrounding the Au atoms. The structure was optimized using Gaussian 16 and Avogadro software and analysed using VESTA software (green - Cl, dark blue - F, light blue - N, white - H, brown - C, yellow - Au): (a) crystal structure of ImEtAuCl$_2$CF$_3$ optimized at B3LYP/Def2TZVPP level of theory with only Au, Cl and C atoms and characteristic bond lengths of Au – C bond from 0 – 2.065Å and C – C between 0 – 1.358Å forming a carbonaceous matrix that surrounds the Au and Cl atoms in a square matrix; (b) the same type of matrix with similar bond lengths of Au – C and C – C with the addition of C – N and C – F bonds to the carbonaceous matrix; no bond between F – C is observes, the F atom placing itself in the middle of the matrix next to the Au atom; H bond and atoms were removed for clarity; (c) YXZ view of the crystal with focus on the fluorine hydrogenated bonds between the molecules; (d) XYZ view with focus on the carbon matrix of the crystal; (e) ZXY view of the crystal lattice with focus on the carbon rings and the fluorine hydrogenated bonds, the Cl$^-$ atoms and N atoms gather around the Au atoms creating a basket while the F$^-$ atoms are involved in creating a bond at 90deg with the next molecules in a homogenous structure. The crystal lattice was built using 2 x 2 x 2 (a x b x c) Miller indices with a lattice unit cell of 4.275Å on X and Y and 80Å on Z at an angle of 60deg on z-axis and 90deg on x- and y-axis. The crystal matrix on all the planes has an inhomogeneous structure, the exceptions being the YXZ at 20deg, XYZ at -10deg to -70deg and ZXY at -10deg planes, where a coherent pattern can be seen between all molecules. Characteristic to the Au fluorinated carbonaceous matrix is the presence of the Au atom in the centre of a carbon ring bonded with a fluorine hydrogenated carbonaceous bond to other layer molecules. The chlorine and nitrogen atoms follow closely the Au through very strong interatomic forces. The lattice allows the presence of four molecules into a unit cell at an 60deg plane angle.

**Single-crystal XRD structure.** 10 mg of 4,5-dichloro 1,3-diethyl imidazolylidene trifluoromethyl gold(I) complex was dissolved in dichloromethane (20 mg/mL), followed by slow addition of pentane/hexane (dichloromethane: Pentane/hexane is 5:1 vol/vol) onto the dichloromethane layer. The layered solutions were kept in dark for crystallization over two weeks at room temperature. Plate shaped crystals of trifluoromethyl gold(I) complex were obtained, hand-picked, and subjected to structure determinations by X-ray diffraction analysis. The crystals belong to the monoclinic space group P2$_{1/c}$ (β = 98.188°, V = 1219.69 Å$^2$) and the unit cell is composed of very weak aurophilic dimer with rather longer Au⋯Au short contact distance (3.772 Å), compared to the previously reported similar dimeric carbene-Au(I)-CF$_3$ complexes. *[5]* The coordination geometry of Au(I) complex is linear (angle C1-Au(I)-CF$_3$ = 176.93°) and slightly distorted, with a tad bit shorter Au-CF$_3$ distance (d$_{Au-CF3}$ = 2.029 Å) compared to previously reported distances found in Au(I) trifluoromethyl complexes containing carbene, isonitrile, or C$_6$F$_5$ ligands (d$_{Au-CF3}$ = 2.031–2.046 Å). *[5]* The molecule has been found to exist in antiparallel conformation with another monomer so as to facilitate closer distance between two Au centres while minimizing steric repulsion between CF₃ and carbene centre, as shown in **Fig 6b**. It is noteworthy to mention that the two methyl groups attached to carbene NCH₂ moiety of a molecule exist in cis configuration to each other, thereby allowing a close approach between two Au centres and widening the distance from the adjacent molecule in next row. Detailed list of bond angle and bond distances have been summarized in **Table 4**.

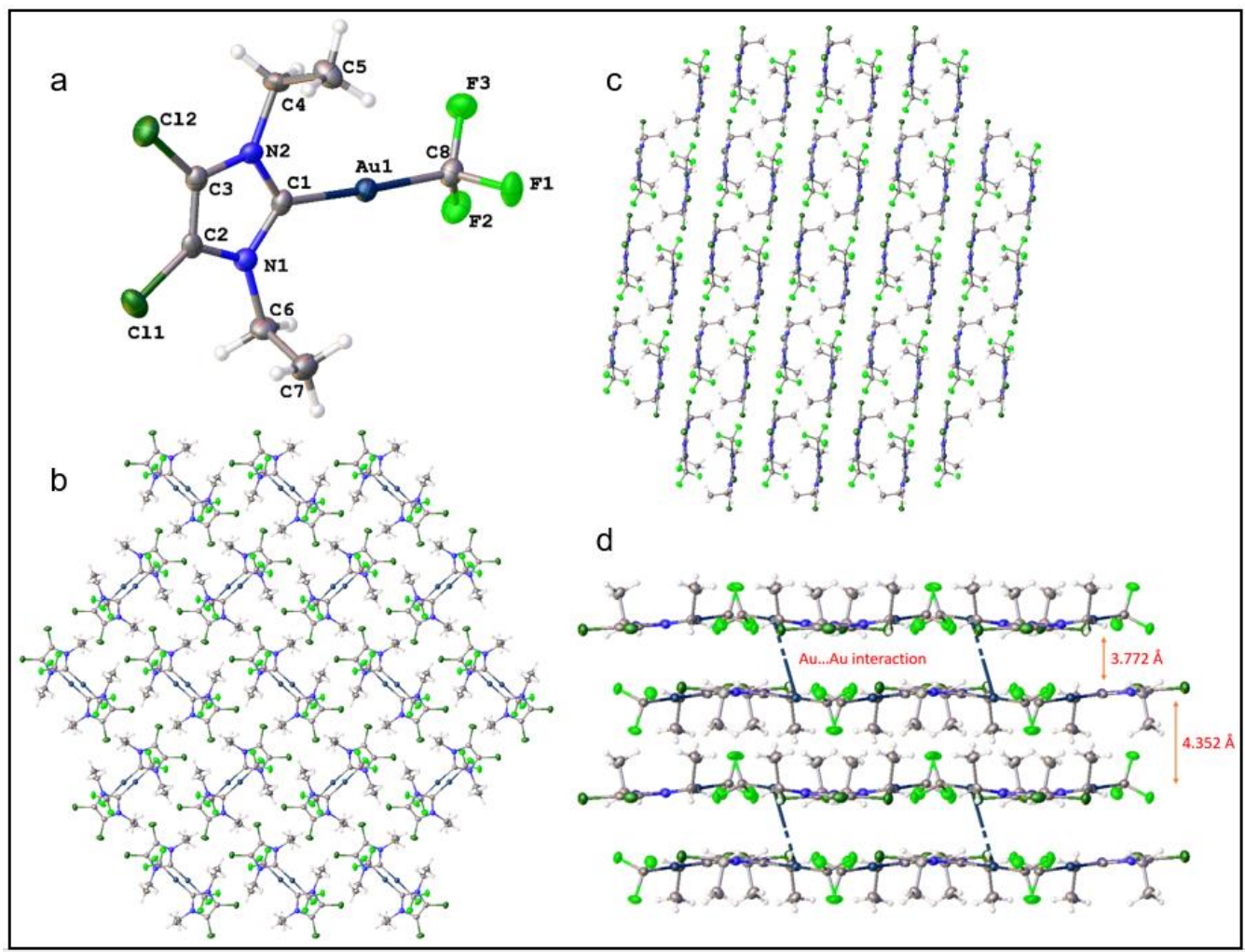

Fig 6: (a) Crystal structure of 4,5-dichloro – 1,3-diethyl – imidazolylidene trifluoromethyl gold(I) complex, ellipsoids have been drawn at 50% probability. Packing and interlayer distance (b) viewed along an axis, (c) *b* axis and (d) the *c* axis has been shown

Detailed investigation of the packing of the crystal structure (**Fig 6b**, **c** and **d**) revealed presence of various weak non-covalent interactions that exist between adjacent rows of molecules. Apart from aurophilic interactions, each molecule engages with four surrounding molecules via four different Van der Wall's interactions which are summarized in **Table 4**. Among these interactions, each molecule engages with two adjacent molecules in the same row via F2···H4A and Cl1···H4B VW interactions and two molecules in the next row via Au1···H5B and H5C···H5C interactions. Interestingly, one Au centre engages with another nearby Au centre via well-known aurophilic interaction (short contact), albeit weak and with a methylene proton of another molecule, positioned in the opposite row to the first one, thereby resulting different interlayer distances. The interlayer distances have been shown in **Fig 6c** where the packing has been viewed along c axis.

| Type of bond | Bond lengths (Å) | Type of angle | Bond angle (°) |
|---|---|---|---|
| Au1 – C8 | 2.029 | C1-Au1-C8 | 176.936 |
| C1 – Au1 | 2.024 | Cl1-C2-C3-Cl2 | 1.166 |
| C1 – N2 | 1.345 | C3-N2-C4-C5 | 88.293 |
| C1 – N1 | 1.348 | C2-N1-C6-C7 | -79.407 |
| C2 – N1 | 1.395 | **Van der Wall's interactions** | |
| C3 – N2 | 1.378 | *Participating atoms* | *Distance (Å)* |
| C3 – C2 | 1.351 | Au1⋯H5B | 3.377 |
| C2 – Cl1 | 1.679 | Cl1⋯H4B | 2.814 |
| C3 – Cl2 | 1.696 | H5C⋯H5C | 2.201 |
| C8 – F1/F2/F3 | 1.373/1.371/1.368 | F2⋯H4A | 2.468 |
| N2 – C4 | 1.469 | **Short contact** | |
| N1 – C6 | 1.463 | Au(I)⋯Au(I) | 3.772 |

Table 4. List of bond lengths, bond angles, torsional angles, Van der Wall's interactions and short contact obtained from single crystal XRD structural analysis

**X-ray powder diffraction data (XRD).** A set of eight experiments at different temperatures and pressure have been run using powder X-ray diffraction method *[71], [72], [73]* to determine the crystallinity and structure of the sample. For the characterization of nanomaterials and deposited complexes at the nanoscale, combinations of tools such as TEM, EXAFS and XRD *[75], [76]* are run to obtain particle size distributions and interlayer plane distances using TEM for the localized nanostructure size and powder XRD for an average nanostructures size. Further measurements can be done for structural characterization of the as deposited nanomaterials using synchrotron radiation (XANES, XSAS e.g.) *[76], [77]*.

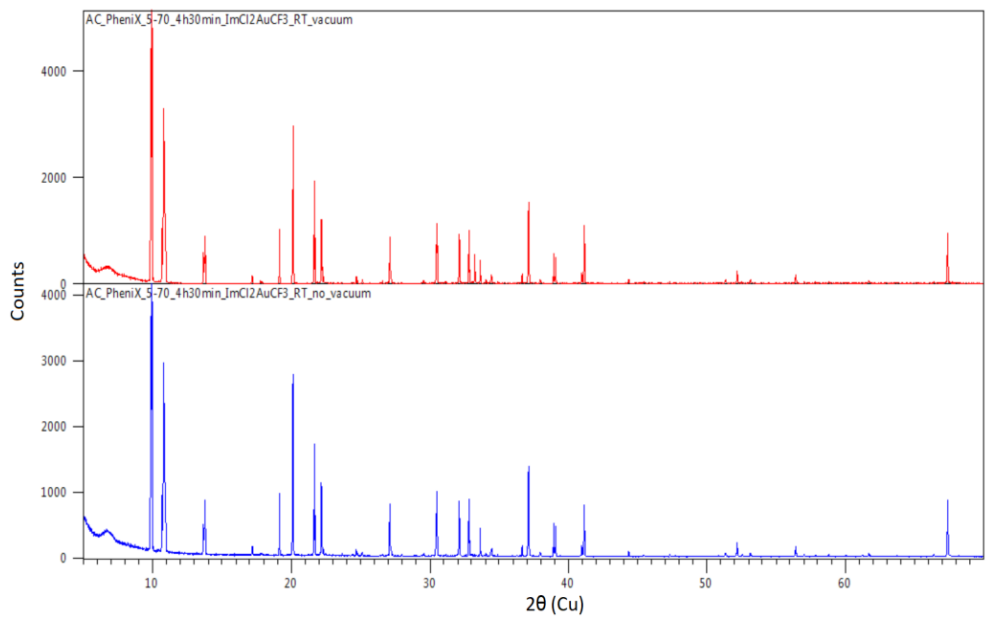

Fig 7. XRD of the Au compound under vacuum condition and at room temperature and atmospheric pressure. The two graphs show very small differences under the two different conditions, highest changes are observed between 30 - 35 2θ(°).

The size of the 4,5-dichloro – 1,3-diethyl – imidazolylidene trifluoromethyl gold(I) grains (**Fig. 8**) is calculated with the formula (1.1), where λ = 1,54060Å, β is the FWHM width of the peak in radians, and θ position of the peaks/2 in radians:

$$D_p = 0.9*\lambda / \beta \cos\theta \qquad (1.1)$$

The highest grain size batch is of 83nm at 0.64rad θ, with an average grain size of all peaks batch of 41nm. Further analysis of the XRD powder data of the compound gives information on the crystallinity by using the position of the 2θ peaks on the XRD spectrum, the spectrum recorded under vacuum was used as cleaner and without $H_2O$ presence. The crystallinity of the precursor is calculated using the equation (1.2) with a value of 54.427%:

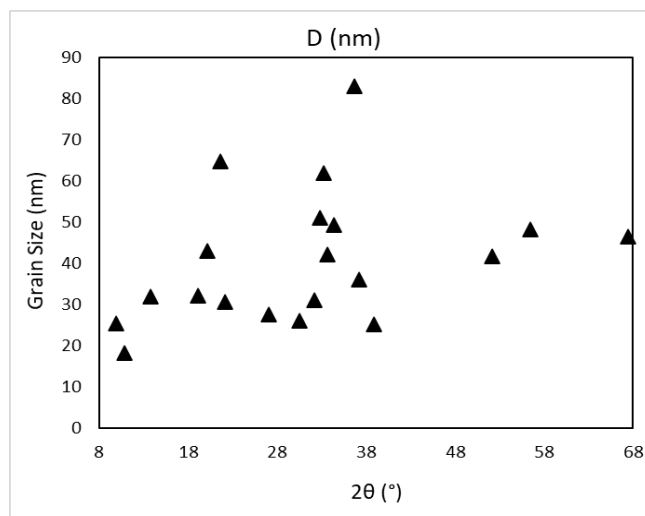

Fig 8. Grain size of Au-compound

$$\text{Crystallinity} = \text{Area of crystalline peaks} * 100 / \text{Area of all peaks (crystalline + amorphous)} \qquad (1.2)$$

A crystallinity of up to 55% is expected, with the grain size limited to an average of 41nm, and the highest grain of 82.86nm, rather small compared to a grown crystal structure or multiple grown crystals in the powder structure, the complex is in amorphous phase mixed with small grains in the form of nanostructures.

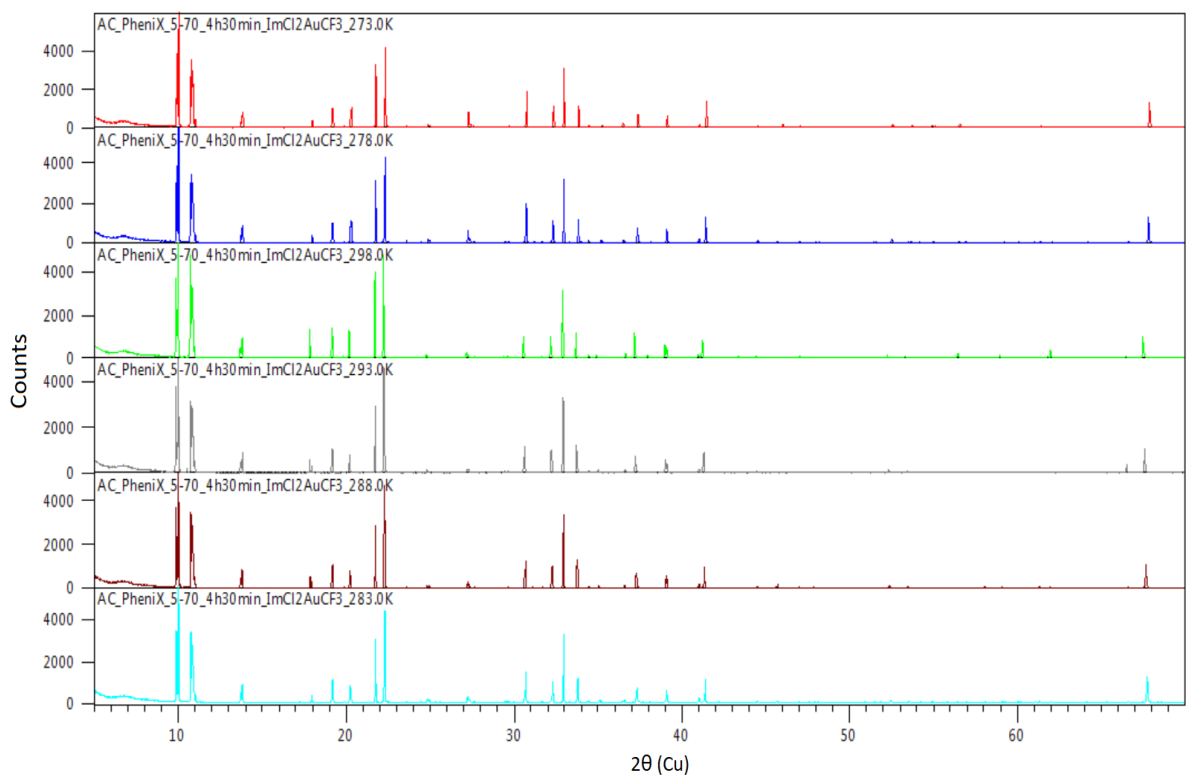

Fig 9. Temperature dependence of XRD data of 4,5-dichloro – 1,3-diethyl – imidazolylidene trifluoromethyl gold(I) complex

An increase in the peak amplitude is observed for the 298K spectrum compared to the rest, a behaviour expected at RT, while separate increase in singular peak amplitude is observed for 273K and 278K at 2θ(°) 26 corresponding to C (002) phase, 298K at 2θ(°) 62 corresponding to Au - chloride (220) BCC plane of the crystalline powder and 293K at 2θ(°) 66.5 corresponding to (400) BCC Au – chloride plane and phase (**Fig. 9**). A separate view of the planes and phases of the crystalline powder and temperature is presented in **Fig. 10**, where phases of BCC are combined with FCC and an intermediate highly hydrogenated layer creating the interspacing of the atom - network positioned between FCC and BCC is observed, the Au surrounding itself with chlorine atoms.

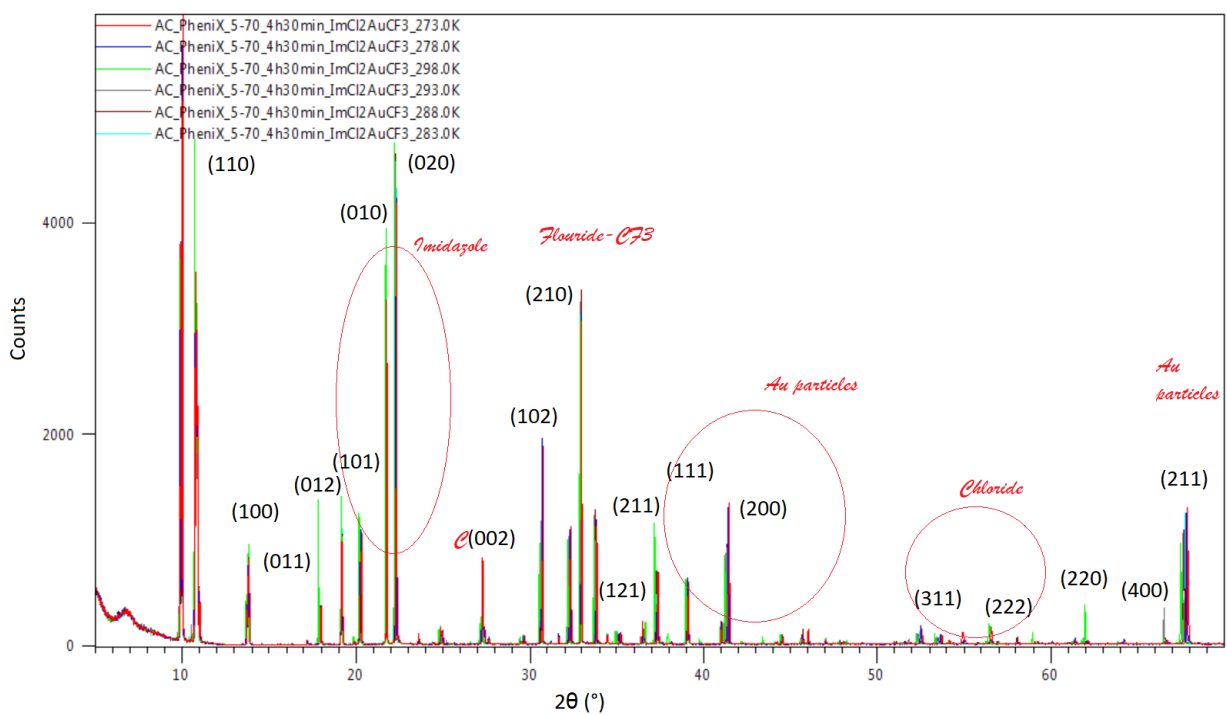

Fig 10. Temperature dependence of the XRD data of the Au compound between 37.5 – 43.5 2θ(°). A normalized behaviour with increasing amplitude and constant delay is observed with changing the temperature from 273 – 298K

**Suitability for FEBID of the precursor**

**Negative ions of 4,5-dichloro – 1,3-diethyl – imidazolylidene trifluoromethyl gold(i).** A number of six gold(i) containing ions are the results of the fragmentation of 4,5-dichloro – 1,3-diethyl – imidazolylidene trifluoromethyl gold(i), as the only anions that contain a metal atom, found in the dissociative electron attachment process of the compound. The two higher mass fragments, $C_5H_{10}N_2F_2AuCl_2^-$ and $C_7H_{10}N_2FAuCl_2^-$, are rather noisy with low cross sections, with highest peak values lower than 1-2cnts. The $C_5H_{10}N_2F_2AuCl_2^-$ ion with a mass of m/z 403amu has a maximum peak at 7.7eV, while the higher mass fragment (m/z 408amu) $C_7H_{10}N_2FAuCl_2^-$ presents three resonances peaking at 0.86eV, 7.4eV and 11.8eV, all with widths in the range of

~1eV. Lower mass fragments, $C_7H_{10}N_2AuCl_2^-$ and $C_5H_9NFAuCl^-$, both have two resonances, with the first resonance peaking before 1eV and the second one after 5eV. With higher cross-section value, the $C_7H_{10}N_2AuCl_2^-$ anion has an average value of count of 150cnts for the maximum resonance falling at an energy of 0.88eV with a width of the resonance peak of 2.03eV, while the second resonance peak is found at an incident electron energy of 7.3eV. The smoothness of the resonance shape comes with the high value of the cross-section as well. A lower value of the cross-section with maximum counts of 2cnts is characteristic to the anionic fragment $C_5H_9NFAuCl^-$ (m/z 334amu) but presenting similar shape to the higher mass ion at m/z 389amu. The electron energy characteristic to the two resonances of the ion (m/z 334amu) are 0.84eV (with a width of 1.81eV) and 5.6eV (with a width of 7.2eV).

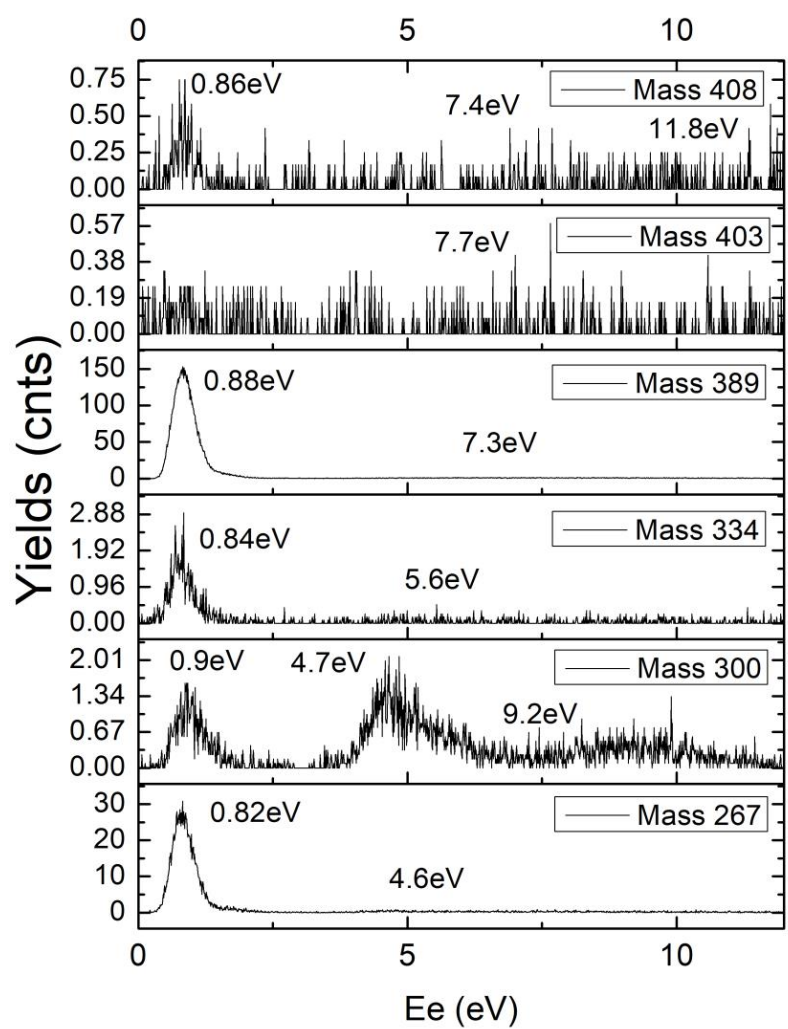

Fig 11. Anions of 4,5-dichloro – 1,3-diethyl – imidazolylidene trifluoromethyl gold(i) m/z 267 to m/z 408

At m/z 300amu an anion without nitrogen or fluoride in its composition is found, though characterized by very low cross-section value. The three peak resonances of the $C_5H_8AuCl^-$ anion are at electron energies of 0.9eV, 4.7eV and 9.2eV, with high characteristic widths and noisier shape. The widths of the three resonances are in the range of ~2-3eV, with values of 1.53eV (0.9eV), 3.38eV (4.7eV) and 3.3eV (9.2eV). The smaller mass anion of the six gold(i) containing ions is $H_4N_2F_2Au^-$ characterized by the presence of nitrogen and fluoride atoms in its composition, and high cross-section. The peak of the highest resonance is found at 0.82eV with a maximum of the peak of 32cnts and a width of 1.67eV. The smaller amplitude resonance falls at 4.6eV having a width of 8eV and a number of counts lower than 2cnts. After the metal containing anions, the precursor fragments in negatively charged organic fragments. The $Cl^-$ ion corresponding to a m/z of 35 presents a resonance peak at 0.85eV with high cross-section values, making it the most abundant anion result of the fragmentation of the precursor at 70eV. A shoulder corresponding to the same peak is observed at 1.8eV, while a wider resonance peak with a width of almost 4eV is observed at 5.3eV. Another very abundant anion is the $C_2H_6NCl^-$ ion having only one resonance peak at 0.84eV with a width of 1.7eV and 24 counts and relative high cross-section compared to the rest of the ions that present values under 10 counts at the height of their peak.

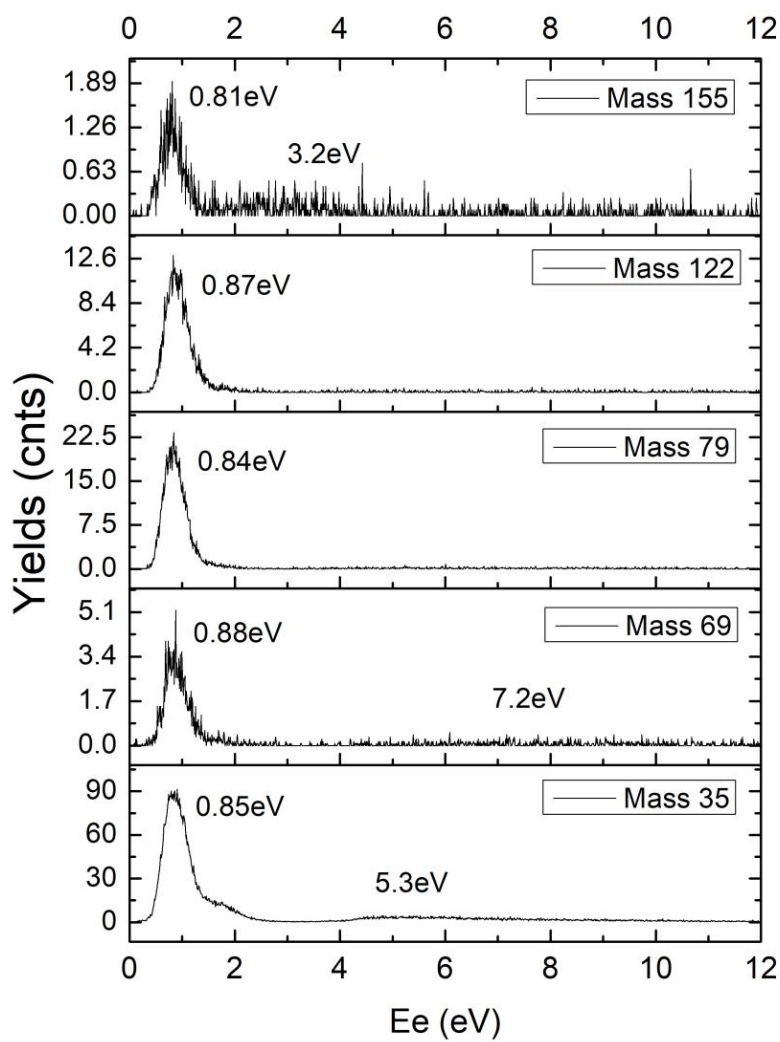

Figure 12. Organic parts and volatile fragment anions of 4,5-dichloro – 1,3-diethyl – imidazolylidene trifluoromethyl gold(i)

The fragmentation of 4,5-dichloro – 1,3-diethyl – imidazolylidene trifluoromethyl gold(i) follows the steps in relation (1) with the result of chloride anion and of a higher mass neutral fragment $C_8H_{10}Cl_2N_2AuF_3$ + e$^-$ →$C_8H_{10}ClN_2AuF_3$ + Cl$^-$ (1). *Xuan et al [12]* presents the fragmentation of the 1,2-dichlorobenzene at low energy DEA studies using ion mass spectroscopy, time-of-flight ions and VMI on the fragmentation of the compound and resultant negative ion of Cl$^-$ at two resonances, 1.2eV and 6eV, the latter being a wider resonance of the anion possibly corresponding to the two isotopes of chloride $^{35}$Cl$^-$ and $^{37}$Cl$^-$. Chloride anion is not an atypical ion in the fragmentation of the compounds containing Cl, the majority of them forming the chloride anion as a product of reaction in the induced chemistry at the interaction of the molecule with electrons. Typical to the fragmentation of diatomic molecules, the 1,2-dichlorobenzene (1,2 - $C_6H_4Cl_2$) undergoes a transition from σ to σ* that further initiates the fragmentation of the molecule with the resulting fragments of $C_6H_4$ - Cl + Cl$^-$; and the chloride anion in the $^1\Sigma_g$* excited state and $O_h$ geometry. A study of four chlorine containing compounds ($CCl_4$, $CH_2Cl_2$, $CH_3Cl$ and $CHCl_3$) *[13]* was presented to DEA fragmentation, each of them exhibiting the presence of the chloride anion at energies close to 0eV. The positions of the resonances of the four anions are presented in **Table 5**. Each of the ions has the highest amplitude peak close to 0eV at an electron energy of 0.0 ± 0.05eV and the second resonance peak between 6eV and 8eV. The Cl$^-$ from $CCl_4$ has a shoulder of the first resonance of chloride falling at 0.75 ± 0.05eV, with a value of the bond dissociation energy of C – Cl ligand of 3.3 ± 0.3eV and a characteristic electron affinity of EA(Cl$_2$) = 2.35 ± 0.1eV; the electron affinity reported by [NIST Database](#) is in good agreement with the values reported by *Scheunemann et al [13]* of 2.5 ± 0.2eV. The two lower in value cross-section fragments, $CF_3^-$ and $C_4H_9N_2Cl_2^-$ lacking the presence of any metal atom, have amplitudes in the range of ~2 – 5 counts, with the highest amplitude peak for $C_4H_9N_2Cl_2^-$ falling at 0.81eV having a width of 1.26eV, while the second peak of the same resonance has its maximum at 3.2eV with the width of the peak of 2.16eV. The $CF_3^-$ anion has its maximum amplitude of the first resonance peak at 0.88eV with a width of 1.5eV and the second peak maximum at 7.2eV characterized by a width of 6.6eV. These four lower mass fragments are particularly interesting through the lack of any metal atom in their composition, depositing and releasing as a result of collision and ionization in the interaction with a secondary electron only volatile fragments and organic material, increasing the level of contamination of the FEBID structures.

*Manaa (2017)* in *[62]* defines the value of the calculated electron affinity from Gaussian 4 simulations at CCSD(T) level of theory as the sum of the values of the energy $E_e$ for the neutral and the anion with added zero-point corrections of the two values (2): EA = [$E_e$(optimized neutral) + ZPE(neutral)] – [$E_e$(anion) + ZPE(anion)]. A similar relation is used for the cation ionization potential at ZPE (3): IP = [$E_e$(cation) + ZPE(cation)] – [$E_e$(optimized neutral) + ZPE(optimized neutral)]. In the DEA Gaussian calculations run at DFT level, the values of the transitions from the HOMO to the LUMO orbitals are related to excitation energies with the formation of a temporary negative ion (TNI). The electron affinity and electronegative potential (absolute electronegativity or absolute hardness) of the anions results of the DEA process of 4,5-dichloro – 1,3-diethyl – imidazolylidene trifluoromethyl gold(i) follow (2) with resulting values in the range of 2.9 – 3.9eV (see **Table 5**). The electron affinities of the products have been calculated at room temperature (298.15K), the zero-point energy corrections being highly sensitive to the input temperature. The highest value of the electron affinity is obtained for mass m/z 267 corresponding to $H_4N_2F_2Au^-$ with an EA value of 3.90094eV (~0.143357 Hartree). Lower electron affinity values are calculated for $Cl^-$ ion and $C_7H_{10}N_2AuCl_2^-$ with values of 3.315709eV (~0.12185 Hartree) and 3.03271eV (~0.11145 Hartree), respectively. The VEA (vertical electron affinities) values from our Gaussian 16 simulations at B3LYP/LANL2DZ level of theory are calculated using the natural bond orbital populations (NBO) and pole p3+ calculations are presented for reference in **Table 5**. The vertical electron affinities (VEA) [67], [68] are calculated for each ion with high cross-sections values ranging from 0.506eV to 1.090eV. The VEA of the 4,5-dichloro – 1,3-diethyl – imidazolylidene trifluoromethyl gold(i) has a value of 0.242eV, 0.146eV higher than the excited anion parent with a value of the VEA of 0.096eV; similar to other compounds containing C - H bonds; example of $CH_3^-$, $SiH_3^-$, and $CHCH_2^-$ in [69]. The anion parent was not determined experimentally as being present due to a short-lived life and being unstable in the $^2A_1'$ state, characteristic to the form of a temporary negative ion (TNI).

| Ions | $C_7H_{10}N_2AuCl_2^-$ | $H_4N_2F_2Au^-$ | $C_2H_6NCl^-$ | $CF_3^-$ |
|---|---|---|---|---|
| **VEA(eV)** | 0.506 | 1.090 | 0.561 | 2.820 [70] *VDE value |
| **EA(eV)** | 3.0327 | 3.9009 | 2.9781 | 3.3157 |

Table 5. Electron affinities of ions $C_7H_{10}N_2AuCl_2^-$(m/z 389), $H_4N_2F_2Au^-$(m/z 279), $C_2H_6NCl^-$ (m/z 79) and $CF_3^-$ (m/z 35)

In the fragmentation of the 4,5-dichloro – 1,3-diethyl – imidazolylidene trifluoromethyl gold(i) two pathways are possible, described by the relations (4) and (5), one of the fragmentation paths results in the formation of $CF_3^-$ (m/z 69) anion with $C_7H_{10}Cl_2N_2Au$ as neutral fragment, while the second fragmentation pathway results in the formation of $C_7H_{10}N_2Cl_2Au^-$ (m/z 389) anion and $CF_3$ as a neutral fragment: $C_8H_{10}N_2Cl_2AuF_3 + e^- \rightarrow C_7H_{10}N_2Cl_2Au^- + CF_3$ (4) and $C_8H_{10}N_2Cl_2AuF_3 + e^- \rightarrow C_7H_{10}N_2Cl_2Au + CF_3^-$ (5). While the $CF_3$ in neutral state is in its ground state $^2A_1$ $C_{3v}$, for the $CF_3^-$ anion the symmetry point group conserves ($C_{3v}$), but the excited state of the anion transitions to $^1A_1$ presents two peaks of the resonance, at 0.88eV and 7.2eV; other higher excited states correspond to $E''_1$ and $E''_2$. Similar behaviour of the $CF_3^-$ anion to $CF_3^-$ from 4,5-dichloro – 1,3-diethyl – imidazolylidene trifluoromethyl gold(i) is observed for the 5-trifluoromethanesulfonyl-uracil [16] compound used extensively in cancer therapy as a radiosensitizer reducing the amount of radiation needed for the treatment of the cancer cells. Defined by an electron affinity of EA = 1.69eV, the fragmentation of 5-trifluoromethanesulfonyl-uracil (OTfU) [16] in a neutral fragment and $CF_3^-$ takes place at an electron energy characterized by four resonance peaks at 0eV, 2.35eV, 4.75eV and 8.45eV with the dissociation of the S – $CF_3$ group ligand. The $CF_3^-$ ion in the hexafluoroacetone azine $((CF_3)_2C = NN = C(CF_3)_2)$ [17] reaction has its resonances falling at higher energies representing typically a bond cleavage with the fragmentation of a C – $CF_3^-$, a C bond to a $CF_3$ having a lower bond dissociation energy than the Au –

CF$_3$ bond cleavage for 4,5-dichloro – 1,3-diethyl – imidazolylidene trifluoromethyl gold(i). Values of 1.61eV are reported for the affinity of the CF$_3^-$ in *[17]* with the DEA resonance peaking at two energies, 3.8eV with the highest amplitude and 7.3eV resonance with lower amplitude, in the range of <10 counts. Calculations of the vertical electron affinities (VEA) and bond dissociation energies show a higher bond strength of the Au – CF$_3$ ligand compared to the organic ligands for the ImCl$_2$EtCF$_3$Au.

**Proton NMR stability measurements.** The proton NMR measurements and water NMR measurements are rather simple measurements used often in the pharmacology industry for stability analysis of complex drugs and compounds and degradation study of these compounds in specified conditions (pressure, temperature, luminosity). The applications of the proton NMR studies are not limited to only stability analysis but have implications to the study of the structure and bonding of complex compounds to certain proteins (RNA signal assignment and validation *[55]*, probing metallic-aromaticity *[53]*) or to structural changes (n-membrane lactones isomerism *[54]*). The most common example of the use of this type of measurement is the proton NMR studies to proteins *[52]* in different storage, transportation and daily-use conditions. The proton NMR and water NMR offer in these circumstances a comprehensive view on the stability of the compound and the time it takes for the chemical complex to degrade or to form new bonds as a result of transition processes to a new form or a chemical reaction induced by temperature or changes in the environmental conditions (pressure or light).

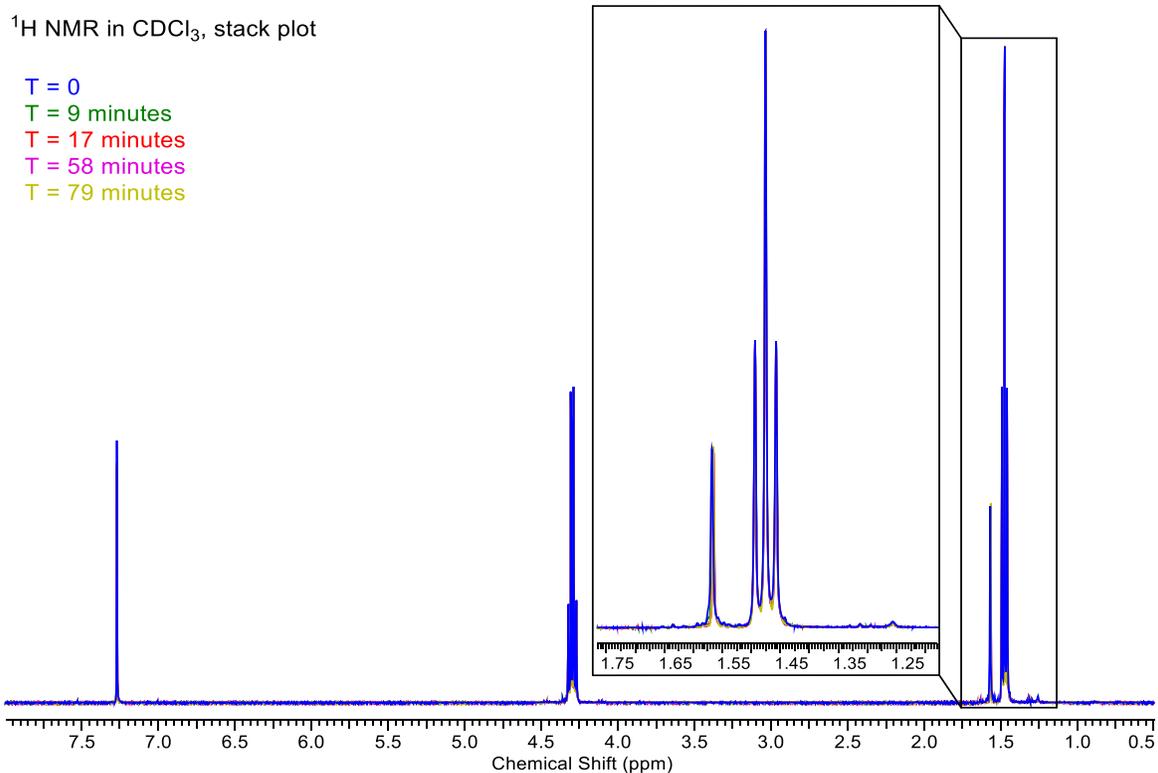

Fig 13. Chemical shift of 4,5-dichloro – 1,3-diethyl – imidazolylidene trifluoromethyl gold(i) in CDCl$_3$ solution

The sample (~5 mg) was dissolved in 1cm$^3$ wet CDCl3 under air; no special precautions were taken other than that the sample was initially transferred into the NMR tube in an Ar glove box. A set of $^1$H and $^{19}$F spectra were queued such that each element was monitored over 80 minutes. $^1$H NMR spectra are referenced to residual protio-solvent. The $^1$H spectra show no significant change over 81 minutes (time in the instrument, ~5 minutes between sample creation and injection into the instrument). There is a slight drift in linewidths, and the resonance at δ1.5 ppm caused by water contamination broadens and shifts from 1.565 to 1.569 ppm, *i.e.,* negligibly consistent with changes in H$_2$O and HCl concentration. The image below shows the stacked spectra and a blown-up portion arising from the CH$_3$ groups.

The $^{19}$F spectra also show minimal change over the time period. The stackplot shows only the first and last spectra recorded, for clarity. The two spectra are essentially identical save for the disappearance of two minor peaks at δ-41.5 ppm and δ-41.8 ppm. In the initial spectrum, these account for ~2% of the total integratable intensity.

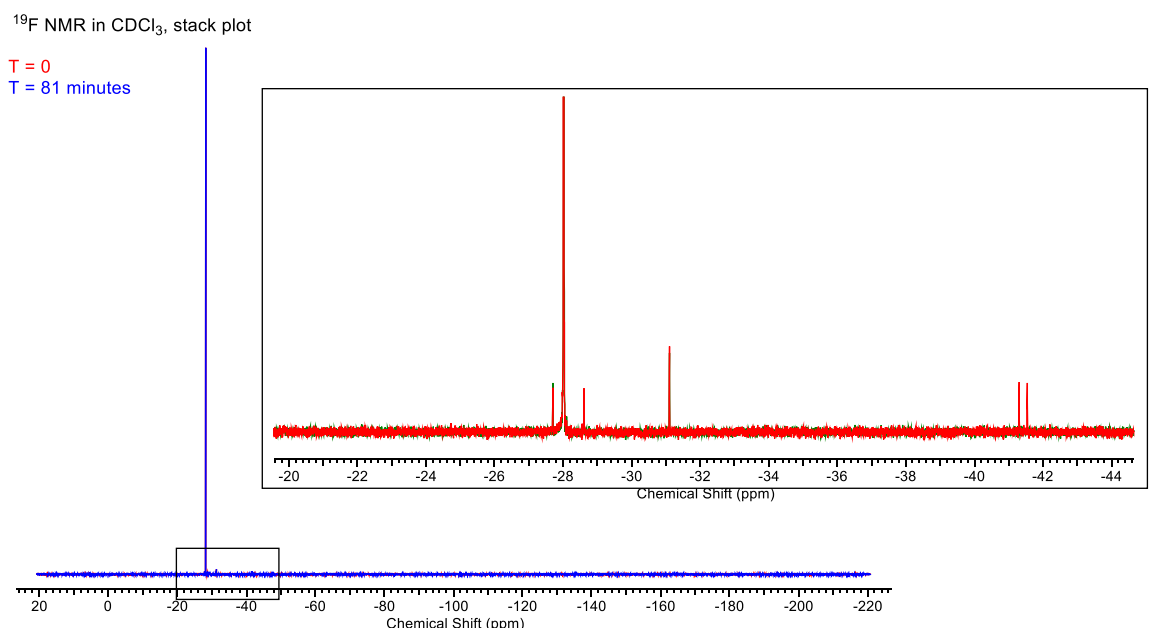

Fig 14. Full spectrum of chemical shift of 4,5-dichloro – 1,3-diethyl – imidazolylidene trifluoromethyl gold(i) in CDCl$_3$ solution

**Gibbs Free Energy of Reaction.** The Gibbs free energy of formation and reaction has been used multiple times to analyse the suitability of different complexes and organic or protein material for drug resistance *[22]*, equilibrium calculations of emulsion systems *[21]*, enthalpy of DNA formation *[41]*, calculations of different minerals formation *[42]* and nanolithography (the present study). The Gibbs free energy of a compound is explained in terms of enthalpy (ΔH), temperature (T) and entropy (S) by the relation: ΔG = ΔH – TΔS. The values of the enthalpy and entropy are taken from simulation data for each of the products and reactants. For a reaction the Gibbs free energy is obtained *[42]* from relation (4) ΔG = G(products) – G(reactants), while for molecular complexes the relation (5) ΔG = G(species) – ∑G(elements) suffices. The Gaussian software works with calculating the corrections to the enthalpy and entropy of formation or reaction based on total energy and contributions from vibrational, rotation, translational or electronic motion; according to this the enthalpy correction is explained by the relation (6) $H_{Corr} = E_{tot} + k_BT$ [43], where the $E_{tot}$ is the total energy. In a similar manner, the entropy is defined by $S_{tot} = S_t + S_e + S_v + S_e$ (7).

Materials in powder form are usually analysed for moisture content mostly in food industry [57 - 59] and to obtain the dissolution rates of complexes [56] through the calculation of the Gibbs free energy, but the use of Gibbs free energy still remains the means to determine the stability of a compound using simulation obtained values of entropies and enthalpies. In order to analyse the suitability of the $Cl_2ImEtCF_3Au$ precursor, the Gibbs free energy from DFT calculations has been used. More industrial oriented applications to pipeline transport industry (water, gas, oil, steam) [60] are by the analysis of Gibbs free energy of solid-state $CO_2$ in the transport of $CO_2$ in carbon capture and storage (CCS). The Gibbs free energy is the entity that defines the probability of a reaction to take place, the volatility and stability of the compound. The values of $\varepsilon_0$, $\varepsilon_{ZPE}$, $H_{Corr}$ and $G_{Corr}$ are calculated from the thermochemistry of $Cl_2ImEtCF_3Au$ at DFT level using B3LYP with a Def2TZVPP basis set, where $\varepsilon_0$ is the electronic energy, $\varepsilon_{ZPE}$ is the zero-point energy, $H_{Corr}$ is the enthalpy correction and $G_{Corr}$ is the Gibbs free energy correction. The sum of the electronic and enthalpy energy, sum of the electronic and Gibbs free energy and sum of the electronic and zero-point energy are used as $\varepsilon_0 + H_{Corr}$, $\varepsilon_0 + G_{Corr}$ and $\varepsilon_0 + \varepsilon_{ZPE}$. The calculated thermochemistry values are presented in **Table 6** with the values obtained from the Gaussian calculations. The calculations have been done at a temperature of 298.15K and a pressure of $1 \times 10^{-4}$Pa. The reaction with the result of an anion and a neutral fragment formation follows the pathway $Cl_2ImEtCF_3Au^* \rightarrow Cl_2ImEtAu^-$ (m/z 389) + $CF_3$ (m/z 69) for which the reaction energy and enthalpy are calculated to obtain the bond dissociation energy (BDE) and bond dissociation free energy (BDFE) of the reactants into products of reaction.

Calculations of Gibbs free energy at atomistic level with great results to modelling of crystal defects are reported by *Cheng and Ceriotti* [61], though at defect sites the model predicts energies 300% higher than evaluated, and at non-defect sites 10% higher than reported for the evaluated model. The higher defect estimated value of the Gibbs free energy is presented as a result of the anharmonicity at the defect sites with the transition at higher temperatures (>298K). The BDE and BDFE are calculated for the chemical reaction taking into account the $\varepsilon_0$ correction to electronic energy and the enthalpy and Gibbs free energy corrections $H_{Corr}$ and $G_{Corr}$ at 298.15K, obtaining the change in enthalpy (1.4) and Gibbs free energy (1.3) with the reaction.

$\Delta_rG = \sum(\varepsilon_0 + G_{Corr})_{products} - \sum(\varepsilon_0 + G_{Corr})_{reactants}$ (1.3)

$\Delta_rH = \sum(\varepsilon_0 + H_{Corr})_{products} - \sum(\varepsilon_0 + H_{Corr})_{reactants}$ (1.4)

The BDEs and BDFEs can be obtained for the formation reaction of the products of reaction, the anion and the neutral fragment $\Delta_fH$ (1.5) and $\Delta_fG$ (1.6):

$\Delta_fH = \sum\Delta_fH_{products}(298K) - \sum\Delta_fH_{reactants}(298K)$ (1.5)

$\Delta_fG = \Delta_fH(298K) - T * (S(298K)_{parent} - \sum S(298K)_{fragments})$ (1.6)

The results of the calculations are presented in **Table 6** with a reaction BDE of -0.665 (Hartree; -417.294kcal/mol).

|  | Cl$_2$ImEtCF$_3$Au$^-$ | Cl$_2$ImEtAu$^-$ | CF$_3$ |
|---|---|---|---|
| $\varepsilon_0$ | -1776.996028 | -1438.642071 | -337.551026 |
| $\varepsilon_{ZPE}$ | 0.187447 | 0.158332 | 0.012158 |
| $E_{tot}$ | 0.206957 | 0.173440 | 0.015612 |
| $H_{Corr}$ | 0.207901 | 0.174385 | 0.016556 |
| $G_{Corr}$ | 0.133052 | 0.110574 | -0.014568 |
| $\varepsilon_0 + \varepsilon_{ZPE}$ | -1776.808581 | -1438.483739 | -337.538867 |
| $\varepsilon_0 + E_{tot}$ | -1776.789071 | -1438.468631 | -337.535414 |
| $\varepsilon_0 + H_{Corr}$ | -1776.788127 | -1438.467686 | -337.534470 |
| $\varepsilon_0 + G_{Corr}$ | -1776.862976 | -1438.531497 | -337.565594 |
| **Reaction and formation products** | Δ$_r$H | Δ$_r$G | S |
|  | -0.786 | -0.665 | 0.046 |

Table 6. Free Gibbs energy correction, enthalpy correction and zero-point corrections of Cl$_2$ImEtCF$_3$Au$^-$, Cl$_2$ImEtAu$^-$ and CF$_3$, products of formation and products of reaction

The metal – ligand bond between Au(I) and CF$_3$ presents higher BDE for Cl$_2$ImEtCF$_3$Au (189.15kcal/mol) than for Au(I) – CF$_3$ in CF$_3$AuCO of 151.4kcal/mol, Au(I) – Cl in ClAuPMe$_3$ of 77.9kcal/mol or Au(I) – Me in MeAuPMe$_3$ of 43.4kcal/mol reported by *Marashdeh et al [23]*. The value reported from our calculations of -424.47kcal/mol is specific for an exogenic process releasing energy, though is not characterized by high cross-section value for the elimination of the Au – CF$_3$ ligand. Lower Δ$_f$G would mean that the molecule is unstable making it hard to work with and to be transferred from the vial through the gas line inside the vacuum chamber. Values as low as +16.5 kcal/mol for ClAuPF$_3$ have been reported in *[23]* rendering the ClAuPF$_3$ compound as one of the compounds with low vaporization pressure. Not stable in air and at room temperature, the Cl$_2$ImEtCF$_3$Au has similar behaviour to AuCF$_3$CO *[23], [24]* that darkens in the presence of heat and light, a sign of the oxidation process. The Cl$_2$ImEtCF$_3$Au is not to be kept at temperatures higher than 5degC as it spontaneously breaks ligands and degrades, while the presence of air would intensify the process of degradation and oxidation.

**Synthesis of the gold(i) compound.** Gold(i) NHC complexes are a class of compounds that is widely known and studied in chemistry for their versatility, among others in catalysis *[44]*, biomedicine *[45]* and photochemistry *[46]*. The most important characteristic of the NHC ligand is the carbene carbon, which is stabilized by two neighboring nitrogen atoms. *[47]* Due to their popularity, several ways for the synthesis of gold(i) NHC complexes have been reported *[48]*, which makes these systems easily accessible and adaptable to required needs. The gold(i) NHC complex *[38]* investigated in this work was synthesized following a reported literature procedure which is presented in **Fig 15**. Starting from 4,5-chloroimidazole the desired NHC ligand precursor was obtained as a salt in a yield of 94 % through two sequential alkylation reactions using ethyl iodide. *[49]* By reacting the NHC ligand precursor with silver oxide, the respective silver complex was formed *in-situ* which underwent a trans-metalation reaction upon the addition of one equivalent of the gold precursor Au(SMe$_2$)Cl. *[50]* The resulting gold NHC chloride complex was isolated in a yield of 92%. In the last step, the title compound was synthesized through another silver mediated trans-metalation reaction. The active silver species AgCF$_3$ is formed *in situ* from the reaction of silver fluoride with Me$_3$SiCF$_3$ and exchanges the chloro-ligand with a CF$_3$ group when Au(NHC)Cl is added, yielding the desired Au(NHC)CF$_3$ complex as a colourless solid (66 %).*[51]*

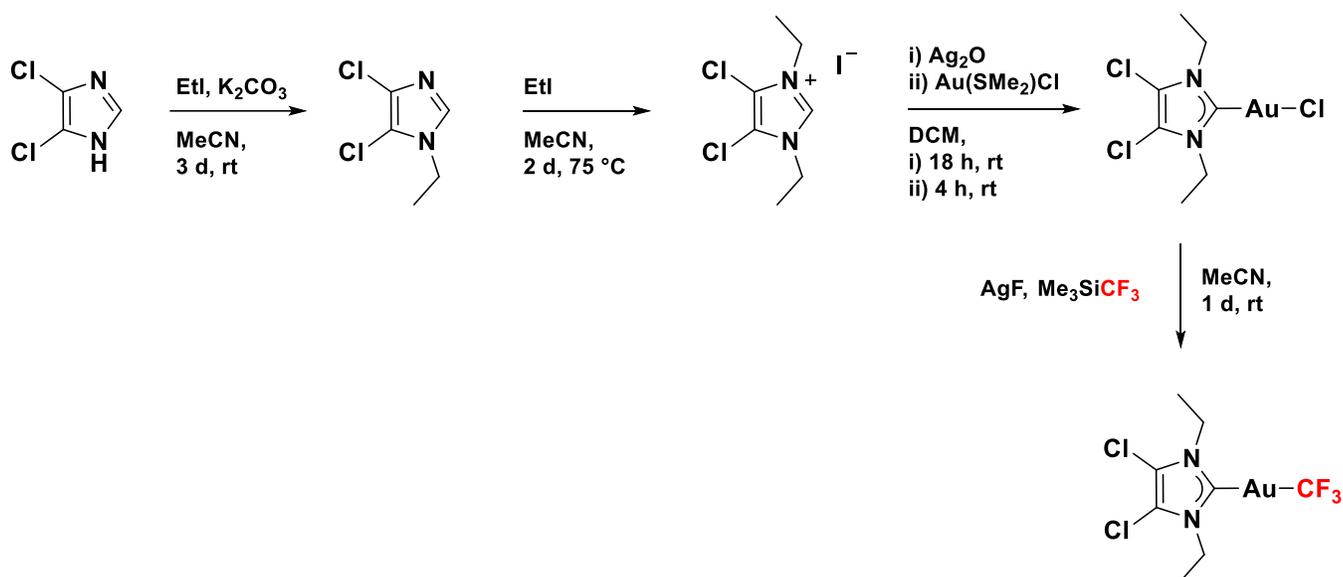

Fig 15. Synthesis route for 4,5-dichloro – 1,3-diethyl – imidazolylidene trifluoromethyl gold(i)

**Computational details.** The simulations of the structure of C$_8$H$_{10}$Cl$_2$N$_2$AuCF$_3$ have been run at DFT/B3LYP level making use of the full orbital populations and natural bond orbitals using a B3LYP/Def2TZVPP basis set. The excited states calculations have been run using TDDFT. The crystal structure and slab for XRD simulations were built using Vesta and Avogadro software.

**Conclusions**

The 4,5-dichloro – 1,3-diethyl – imidazolylidene trifluoromethyl gold(i) was analysed to its suitability as a FEBID precursor. As a newly designed compound specifically for deposition of nanoscale structure, its vaporization pressure, stability in air and volatility have been studied using proton NMR and Gibbs free energy of reaction.

A good volatility value was obtained for the compound and a high stability in air with very low modifications of the structure during exposure. Its fragmentation, resonances and anions at low electron energies and DEA have been obtained using Velocity Map Imaging studies with great success. The structure, packing, orientation of the planes and grain size have been run making use of powder XRD diffractometer data and the VESTA simulation software has offered reliable insights into the crystalline vs amorphous structure of the compound.

**Acknowledgements.** We want to acknowledge MP receiving funding from the European Union's Horizon 2020 research and innovation program under the Marie Skłodowska-Curie grant agreement No 722149, and the work of our partner institutions J. Heyrovský Institute of Physical Chemistry of the Czech Academy of Sciences and University of Oslo.

No conflict of interests has been declared.